\documentclass[prl,twocolumn,aps,showpacs,nofootinbib,floatfix,groupedaddress]{revtex4-1}
\usepackage{amsmath}
\usepackage{amssymb}
\usepackage{graphicx}
\usepackage{verbatim}
\usepackage[colorlinks=true,linkcolor=blue,citecolor=blue,urlcolor=blue]{hyperref}
\usepackage{txfonts}
\usepackage{xr}

\usepackage{pdfpages}
\pdfoutput=1
\usepackage{etoolbox} 

\makeatletter
\patchcmd{\@outputpage@head}{\@ifx{\LS@rot\@undefined}{}{\LS@rot}}{}{}{}
\makeatother

\usepackage{ifthen}
\newcommand{\hideandshow}[1]{%
 \ifthenelse{\isundefined{\showme}}{}{#1}}

\begin{document}

\title{Cavity-mediated electron-photon superconductivity}
\author{Frank Schlawin$^{1}$, Andrea Cavalleri$^{1, 2}$ and Dieter Jaksch$^{1}$}
\address{$^{1}$ Clarendon Laboratory, University of Oxford, Parks Road, Oxford OX1 3PU, United Kingdom}
\email{frank.schlawin@physics.ox.ac.uk}
\affiliation{$^{2}$ Max Planck Institute for the Structure and Dynamics of Matter, Luruper Chaussee 149, 22761 Hamburg, Germany}

\begin{abstract}

We investigate electron paring in a two-dimensional electron system mediated by vacuum fluctuations inside a nanoplasmonic terahertz cavity. 
We show that the structured cavity vacuum can induce long-range attractive interactions between current fluctuations which lead to pairing in generic materials with critical temperatures in the low-Kelvin regime for realistic parameters. The induced state is a pair density wave superconductor which can show a transition from a fully gapped to a partially gapped phase - akin to the pseudogap phase in high-$T_c$ superconductors.
Our findings provide a promising tool for engineering intrinsic electron interactions in two-dimensional materials.
\end{abstract}

\maketitle

Pairing between fermionic quasi-particles in solids through the exchange of virtual bosonic excitations is one of the most studied phenomena in the solid state because it can lead to remarkable emergent phases of matter like superconductivity. This phenomenon has been extensively investigated for the case of attractive interactions induced by phonons \cite{Carbotte90} and magnetic excitations \cite{Lee06}. Vacuum fluctuations of the transverse electromagnetic field can also provide attractive interactions, albeit with far lower efficiency. 
These current-current interactions in the Fermi liquid have to date been discussed in \textit{free} space, where effects are restricted to very low temperatures \cite{Holstein73, Reizer89a, Reizer89b, Khveshchenko93a, Khveshchenko93}, or in strongly correlated materials, where they could lead to the pairing of spinons \cite{Lee07, Galitski08}. 
However, interactions between current fluctuations and vacuum excitations of the electromagnetic field can occur within a cavity over mode volumes far below the free-space diffraction limit $\lambda^3$ (e.g. $V \sim 10^{-5} \lambda^3$ in \cite{Maissen14}). These conditions are routinely created in nanoplasmonic cavities in the THz regime \cite{Zurich12, Liu14, Smolka14, Maissen14, Zhang16, Bayer17, Keller17}, where the ultrastrong coupling regime between light and matter can be reached even in bad cavities. This compression of the cavity field increases the vacuum field strength ($\sim 1/ \sqrt{V}$), and thereby enhances the induced interaction ($\sim 1 / V$) to an experimentally accessible regime. 

Furthermore, the cavity structures the electromagnetic vacuum by inducing a photonic bandgap $\hbar \omega_0$ and creating an effective photon mass $\hbar \omega_0 / c^2$ (where $c$ is the speed of light in the cavity filling material) that reduces its group velocity. 
The extremely low temperature scale of the effect in free space, which is attributed to the smallness of the Fermi velocity with respect to the speed of light \cite{Khveshchenko93a}, should thus be enhanced further.  
In contrast to polariton-mediated superconductivity \cite{Laussy10, Cotlet16, Kavokin16}, where the glue for electron pairing is provided by the exchange of hybridised exciton-photon or phonon-polariton \cite{Sentef18} states in a cavity, the direct interaction with a cavity does not require the pumping of the cavity, and is also not affected by excitonic interactions. Thermal excitations of the cavity field are prevented by sufficiently low temperatures. 

In this letter, we show that the coupling of a THz cavity field to intraband transitions of a two-dimensional electron gas gives rise to a superconducting instability with a critical temperature that can reach the low-Kelvin regime using realistic cavity and material parameters. The electron-photon coupling preserves the electron spin, such that the cavity can mediate both singlet and triplet pairing. 
The realized state is determined by short-ranged electronic interactions in the sample material. 

\begin{figure*}[t]
\centering
\includegraphics[width=0.8\textwidth]{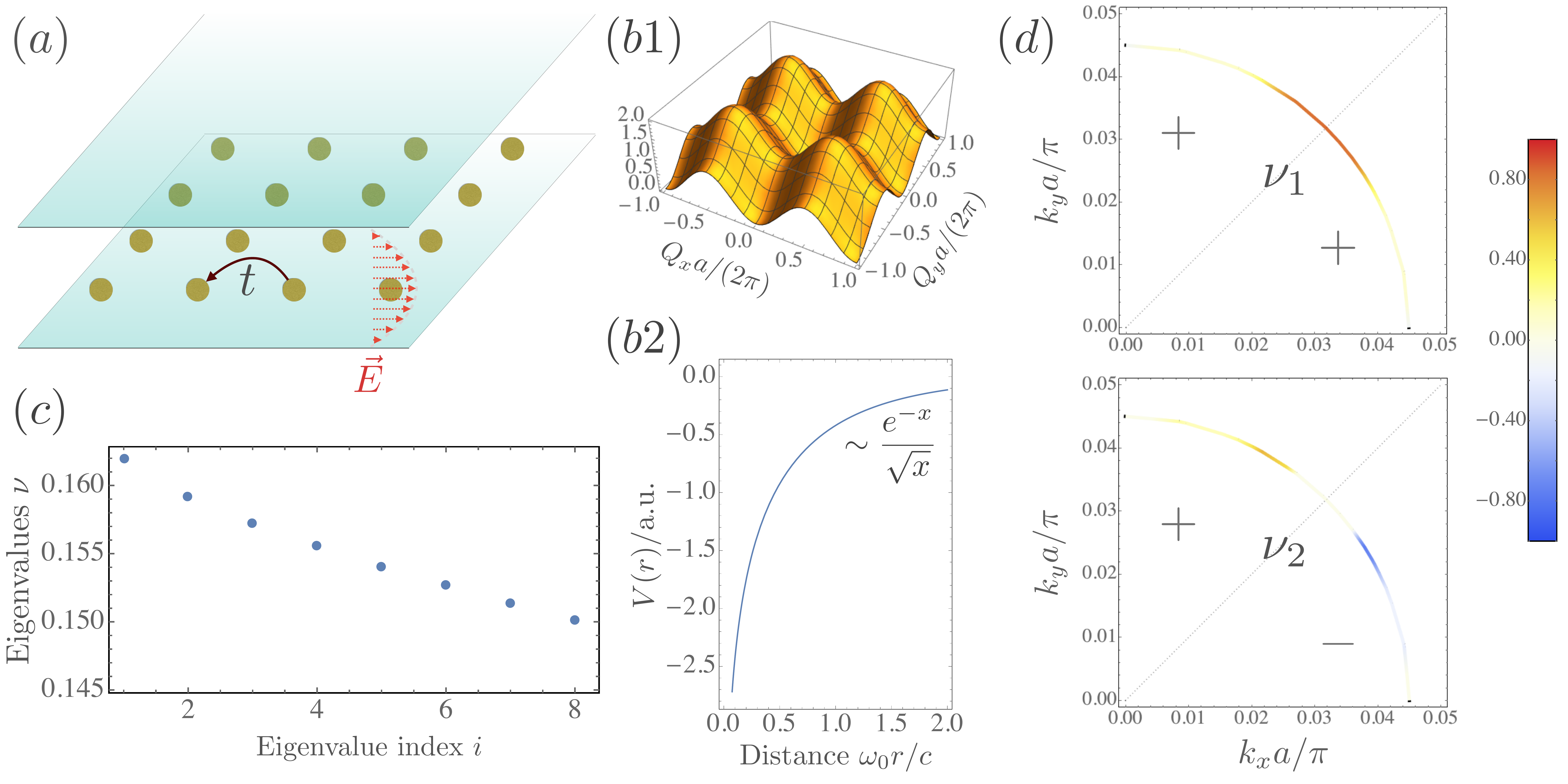}
\caption{
(a) A 2d electron lattice depicted by yellow sites interacts with a cavity, whose electric field distribution is indicated in red. The electronic hopping amplitude $t$ is also sketched.
(b) The exchange of virtual photons between electric intraband transitions creates an effective electron interaction. This is decomposed into two contributions as described in the main text: 
(b1) displays the effect of the lattice geometry on the cavity-mediated interaction across the first Brillouin zone as a function of the nesting vector $\vec{Q}$.
(b2) shows the interactions tailing off as $e^{-x } / \sqrt{ x }$, with $x = \omega_0 r /c$, and diverging logarithmically at the origin.
(c) Leading eigenvalues $\nu_i$ according to Eq.~(\ref{eq.matrix}). 
(d) The eigenfunctions corresponding to eigenvalues $\nu_1$ and $\nu_2$ pertaining to a nesting vector $\vec{Q} = k_0 (1, 1)$ are shown along the positive quadrant of the Fermi surface, and positive and negative sectors of wavefunctions are indicated. The colour code is given on the right. $\nu_1$ corresponds to an singlet, and $\nu_2$ to triplet symmetry.
 We used parameters appropriate for a 2D electron gas in GaAs; relative permittivity $\epsilon = 13$, electron band mass $m^\ast = 0.069 m_e$, chemical potential $\mu = - 3.98 t$, lattice constant $a = 5.6$~$\AA$, and a cavity frequency $\omega_0 = 2\pi \times 5$~THz.
}
\label{fig.setup}
\end{figure*}

We consider a two-dimensional electron gas placed into a cavity made of two mirrors [see Fig.~\ref{fig.setup}(a)].
The electronic Hamiltonian is given by $H_0 =  \sum_{\vec{k},\sigma} \epsilon_{\vec{k}} c^{\dagger}_{\vec{k} \sigma} c_{\vec{k} \sigma}$, where $c_{\vec{k} \sigma}$ destroys an electron with quasi-momentum $\vec{k}$ and polarization $\sigma$, and the electron dispersion $\epsilon_{\vec{k}}$ is determined by the lattice geometry. The multimode cavity is described by the Hamiltonian $H_f = \sum_{\vec{q},s} \hbar \omega_q a^{\dagger}_{\vec{q},s} a_{\vec{q},s}$, where $a_{\vec{q}, s}$ describes the annihilation operator of a photon with in-plane momentum $\vec{q}$ and polarization $s$, and with dispersion $\omega_q = \omega_0 (1 + (c |\vec{q}| / \omega_0)^2)^{1/2}$. 
To leading order, the interaction between cavity and electron system is given by the paramagnetic coupling of the form \cite{Fox01, Dressel02}
\begin{align}
H_{\text{int}}  &= \sum_{\vec{k}, \sigma, \vec{q}, s} \frac{  g_{\vec{k}, s}^{(\vec{q})}  }{\sqrt{N}} \left( a_{\vec{q}, s} + a^{\dagger}_{- \vec{q}, s}   \right) c^{\dagger}_{\vec{k} + \vec{q}, \sigma} c_{\vec{k}, \sigma}, \label{eq.H_int}
\end{align}
where $g_{\vec{k}, s}^{(\vec{q})}$ is a coupling parameter, and $N$ indicates the number of sites. We model the cavity mode compression below the diffraction limit by a compression factor $A$, such that $g \propto 1 / \sqrt{A}$ (see SI).
Note that we have not invoked the commonly used dipole approximation, nor the rotating wave approximation which can fail in the strong-coupling regime in a two-dimensional electron gas \cite{Li18}. While our calculations are based on a specific model for the cavity geometry shown in Fig.~\ref{fig.setup}(a) (details can be found in the supplementary information (SI)), we expect the main results of our paper to be independent of the specific details of the cavity mode.

Integrating out the cavity degrees of freedom, which are in their ground state, and using that $\hbar \omega_q \gg \vert \epsilon_{\vec{k}} - \epsilon_{\vec{k}'} \vert$, the interaction Hamiltonian~(\ref{eq.H_int}) yields the cavity-mediated electron interaction
\begin{align}
V_{\text{eff}} &= - \frac{1}{N} \sum_{\vec{k}, \sigma, \vec{k}', \sigma'} \sum_{\vec{q}, s} \frac{ g_{\vec{k},s}^{(\vec{q})} g_{\vec{k}', s}^{(- \vec{q})}}{\hbar \omega_{q}}  c^{\dagger}_{\vec{k} + \vec{q}, \sigma} c_{\vec{k}, \sigma} c^{\dagger}_{\vec{k}' - \vec{q}, \sigma'} c_{\vec{k}', \sigma'}.  \label{eq.H_eff}
\end{align}
Eq.~(\ref{eq.H_eff}) describes the interaction between current fluctuations mediated by the structured electromagnetic vacuum.  
Due to the symmetry of the electromagnetic field, this interaction is attractive for electrons propagating in the same direction. This ``Amperean" pairing instability can give rise to superconductivity with the Cooper pairs residing on the same side of the Fermi surface, and having a large center-of-mass momentum $\sim 2 k_f$, where $k_f$ is the Fermi wavevector \cite{Lee07, Galitski08}. The resulting pair density superconductor has been discussed in the pseudogap phase of cuprate superconductors \cite{Lee14} and in topological insulators \cite{Kargarian16}, yet it has never been identified unambiguously to the best of our knowledge. 

Amperean pairing can lead to the condensation of electron pairs in the vicinity of a nesting vector $\vec{Q}$ on the Fermi surface \cite{Lee07}. Relabeling the indices, $\vec{k} = \vec{Q} + \vec{p}'$, $\vec{k}' = \vec{Q} - \vec{p}'$ and $\vec{q} = \vec{p}-\vec{p}'$, the cavity-mediated electron interaction~(\ref{eq.H_eff}) reads
$
V_{\text{eff}} = \frac{1}{2 N} \sum_{\vec{Q}, \vec{p}, \vec{p}'} \sum_{\sigma, \sigma'} \! V^{(\vec{Q})}_{\vec{p}, \vec{p}'} \; c^{\dagger}_{\vec{Q} + \vec{p}, \sigma} c^{\dagger}_{\vec{Q}-\vec{p}, \sigma'} c_{\vec{Q}- \vec{p}', \sigma'} c_{\vec{Q}+\vec{p}', \sigma}, \label{eq.V_eff}
$
with $V^{(\vec{Q})}_{\vec{p} \vec{p}'} = - 2 \sum_s g_{\vec{Q}+\vec{p}', s}^{(\vec{p} - \vec{p}')} g_{\vec{Q}-\vec{p}', s}^{(\vec{p}' - \vec{p})} / \hbar \omega_{|\vec{p} - \vec{p}'|}$. 
In the following, we will investigate this attractive interaction on a simple rectangular lattice with nearest-neighbour hopping $t$, for which the electron dispersion reads $\epsilon_{\vec{k}} = - 2t (  \cos (k_x a) + \cos (k_y a) ) - \mu$, where $a$ denotes the lattice constant, and where we also added the chemical potential $\mu$. We find (see SI)
\begin{align}
V^{(\vec{Q})}_{\vec{p}, \vec{p}'} &= - \frac{V_0}{2} \frac{ \sum_{i = x, y} \big( \cos [ (p_i + p'_i) a ] - \cos [2 Q_i a] \big)}{\left( \omega_0 a / c  \right)^2 + \left( (\vec{p} - \vec{p}' )a \right)^2}, \label{eq.V_kk}
\end{align}
where we defined the overall interaction strength $V_0 = 2 g_0^2 (a \omega_0 / c)^2 / \hbar \omega_0$, and $g_0$ denotes the coupling strength which contains the cavity compression. The interaction~(\ref{eq.V_kk}) is attractive, as long as $\cos [(p_i + p'_i) a] > \cos [2 Q_i a]$.
The current operator conserves the electron spin, and as such, it can facilitate electron pairing in both the singlet and the triplet channel.

The potential $V^{(\vec{Q})}_{\vec{p}, \vec{p}'}$ peaks sharply around $\vec{p} = \vec{p}'$, such that it can affect only a very narrow band around the Fermi energy. The width of this peak is determined by the cavity properties encoded in the denominator which only depends on $|\vec{p} - \vec{p}'|$. In the corresponding coordinate space $r$, the potential is proportional to the zeroth-order Bessel function of the second kind, 
\begin{align}
V (r) &\sim - K_0 \left( \frac{\omega_0 r}{c} \right), \label{eq.V2(r)}
\end{align}
shown in Fig.~\ref{fig.setup}(b2). It shows a logarithmic divergence at the origin, and tails decaying as $\sim \exp (- \omega_0 r / c) / ( \omega_0 r / c )^{1/2}$. The exponential decay is a consequence of the off-resonant coupling. 
At larger distances, the virtual photons take on a real character~\cite{Salam}, and the interaction is suppressed by energy conservation. 
However, the length scale of this decay, $c / \omega_0$, is $\sim \mu$m for THz cavities, and thus up to four orders of magnitude larger than typical lattice constants $a \sim 10\; \AA$. For all practical purposes, the cavity-mediated interaction can therefore be considered a long-range interaction.

The numerator of $V^{(\vec{Q})}_{\vec{p}, \vec{p}'}$ depends solely on $(\vec{p}+\vec{p}')$ and the nesting vector $\vec{Q}$. 
For small $\vec{p}$ and $\vec{p}'$, it reduces to $\sim \sin^2 (Q_x a) + \sin^2 (Q_y a) $, which is plotted in Fig.~\ref{fig.setup}(b1). Clearly, the interaction is strongest along the (anti-)diagonals $Q_y = \pm Q_x$. So in the following, we will investigate pairing in the vicinity of the points $\vec{Q} = k_0 (\pm 1, \pm 1)$, where $k_0$ is chosen such that $\vec{Q}$ lies on the Fermi surface. 

We decouple the electron interactions with the mean fields $\Delta^{(\vec{Q})}_{\sigma \sigma'} (\vec{p}) = N^{-1} \sum_{\vec{p}'} V^{(\vec{Q})}_{\vec{p}, \vec{p}'} \langle c_{ \vec{Q} - \vec{p'}, \sigma'} c_{\vec{Q} + \vec{p'}, \sigma} \rangle$. Since $V^{(\vec{Q})}_{\vec{p}, \vec{p}'}$ is attractive only in the vicinity of $\vec{Q}$, we can evaluate pairing around the different nesting vectors separately, and each of these turns into an identical calculation. 
To identify the dominating gap symmetry, we follow the approach in \cite{Scalapino86, Romer15} adopted for the pair density wave state (see SI): We linearise the gap equation near the critical temperature $T_c$, and determine the largest eigenvalues $\nu$ of the equation
\begin{align}
- \frac{1}{2} \int_{\vec{Q}+\vec{p}' \in FS} \frac{d\vec{p}'}{(2\pi)^2} \frac{1}{| \vec{v} (\vec{p}') |} V^{(\vec{Q})}_{\vec{p}, \vec{p}'} \Delta^{(\vec{Q})} (\vec{p}') &= \nu \; \Delta^{(\vec{Q})} (\vec{p}), \label{eq.matrix}
\end{align}
where the line integral runs over vectors $\vec{p}$, such that $\vec{Q} + \vec{p}$ is on the Fermi surface (FS), and $\vec{v} (\vec{p}) = (\nabla_{\vec{Q}+ \vec{p} } \epsilon_{\vec{Q}+ \vec{p} } + \nabla_{\vec{Q}- \vec{p} } \epsilon_{\vec{Q}- \vec{p} }  ) / 2$.
We discretise the quadrant of the FS containing $\vec{Q}$ with a very fine grid, and solve the matrix equation iteratively with the Arnoldi algorithm implementation in Mathematica. The critical temperature is then given by $k_B T_c = 1.13 \hbar \omega_c e^{- 1/\nu} $ \cite{Sigrist}, where $\omega_c$ denotes the cut-off frequency, which we set equal to the basic cavity frequency $\omega_0$.
In the following, we use values appropriate for a two-dimensional electron gas in GaAs heterostructures, keeping in mind that the critical temperature will also be affected by details of the cavity field (see SI). The GaAs system is well described by an effective mass description of the single electrons \cite{Stemmer14}, and phonon scattering is negligible below 1K \cite{Manfra14}.
Using a cavity compression $A = 2\times 10^{-5}$ ($i.e.$ $V \sim 2 \times 10^{-5} \lambda^3$) in Fig.~\ref{fig.setup}(c), we obtain dimensionless coupling strengths $\nu \sim 0.16$. 
This translates into possible critical temperatures reaching into the low-Kelvin regime (see also SI for discussion). 
We have thus established that cavity-mediated electron interactions in electron gases could readily be detected with existing THz cavity technology. Since cavity mode volumes as small as $10^{-10} \lambda^3$ have been reported recently \cite{Keller17} (although at sub-THz frequencies), higher transition temperatures might even be conceivable. The description of this stronger cavity compression would require a strong-coupling extension of our theory, since the corresponding eigenvalues could reach or become larger than unity. In the remainder of this letter, we shall be concerned with the unusual properties and consequences of the fact that the electron pairing here is controlled externally by properties of the cavity.

\begin{figure*}[ht]
\centering
\includegraphics[width=0.9\textwidth]{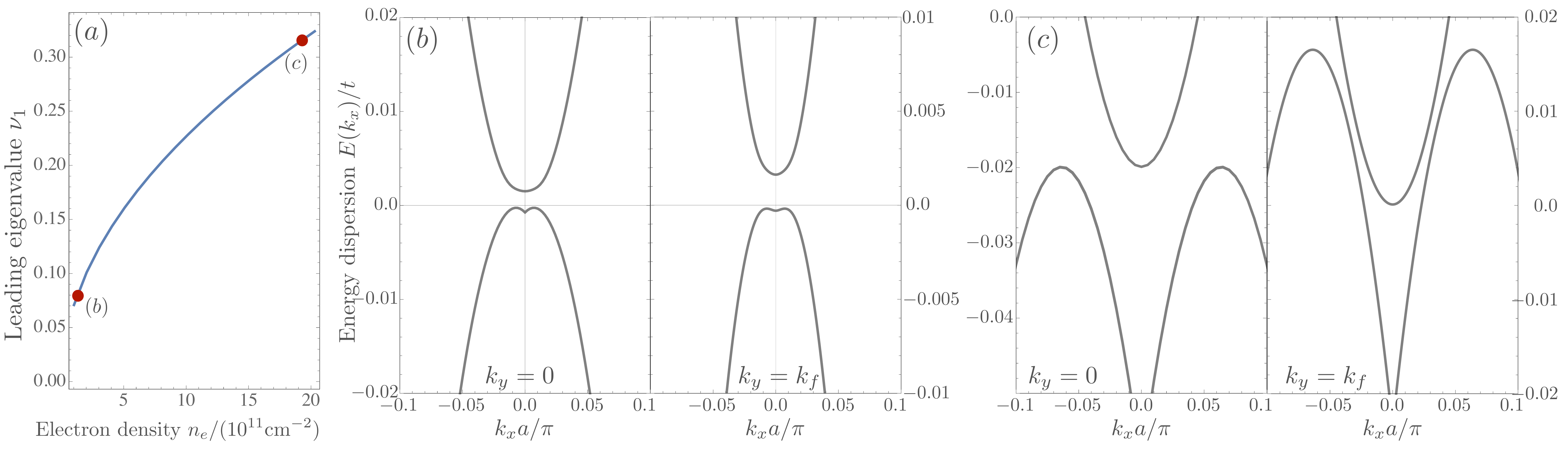}
\caption{
(a) The leading eigenvalue of Eq.~(\ref{eq.matrix}) is plotted as a function of the electron density $n_e$, corresponding to chemical potentials $\mu \in [ -3.999, -3.98 ] t$, and a cavity compression factor $A = 10^{-5}$. 
(b) Electron dispersion $E (k_x)$ along $k_x$, with $k_y = 0$ (left) and $k_y = k_f$ (right) for low electron density $n_e \simeq 0.3 \times 10^{11}$~cm$^{-2}$, and $\Delta_0 = 10^{-3} t$. The axes $E (0)$ and $k_x = 0$ are indicated for orientation.
The system is fully gapped everywhere along the Fermi surface.
(c) the same as (b) with larger electron density $n_e \simeq 2 \times 10^{12}$~cm$^{-2}$, and $\Delta_0 = 10^{-3} t$. The system is gapped only along certain directions in reciprocal space.
}
\label{fig.fermi-arc}
\end{figure*}

The dependence of the pairing strength on the electron density $n_e$ is investigated in Fig.~\ref{fig.fermi-arc}(a), where we plot the leading eigenvalue $\nu_1$ vs. $n_e$ for a fixed cavity compression factor $A = 10^{-5}$ over a large range of densities. An increase of the electron density by a factor $10$ translates into a similar change of $\nu_1$. Hence, the cavity compression factor can always outweigh a change of the electron density, and pairing should be observable even in a low-filling regime with $n_e \lesssim 10^{11}$~cm$^{-2}$ - given a sufficiently strong cavity compression to push the leading eigenvalue to $\sim 0.1$. In the following, we fix the largest eigenvalue at a value corresponding to a zero-temperature mean field value $\Delta_0$, and investigate the gap structure at different electron densities. 
The Amperean pairing around a nesting vector $\vec{Q}$ gives rise to a spatial modulation of the superconducting order parameter in real space, $\Delta (\vec{r}) \sim \cos ( \vec{Q} \cdot \vec{r})$ (assuming equal pairing amplitude on opposite sides of the Fermi surface, $i.e.$ for $\vec{Q}$ and $-\vec{Q}$). The electron density determines $\vec{Q}$, and hence should strongly influence the emerging gap structure.
 
In Fig.~\ref{fig.fermi-arc}(b), the emergent quasiparticle dispersion curves $E (\vec{k})$ (see SI) of the two highest-energy states are plotted vs. $k_x$ along two cuts, $k_y = 0$ and $k_y = k_f$, respectively. Evidently, the dispersion is not symmetric with respect to the Fermi energy, as the particle-hole symmetry is broken in the pair-density wave state, and the hole dispersion is split into two maxima at $\pm Q_x / 2$. Still, at this low electron density, the quasiparticle spectrum is gapped along the whole Fermi surface. 
At larger electron densities, as shown in In Fig.~\ref{fig.fermi-arc}(c), the nesting vector $\vec{Q}$ increases with the Fermi energy, and so does the split in the hole dispersion. As a consequence, the gap is closed along parts of the Fermi surface (e.g. at $k_y = k_f$). This can be understood by reference to Fig.~\ref{fig.setup}(b1): At higher electron densities, the coupling strength becomes more anisotropic. Hence, the condensate is created in the close vicinity of the nesting vectors. Conversely, at low densities the coupling becomes isotropic. The eigenfunction in Fig.~\ref{fig.setup}(d) broadens, and the entire Fermi surface is gapped.

In cuprate superconductors, this pair density wave state is considered as a possible candidate responsible for the pseudogap \cite{Lee14}, where parts of the Fermi surface remain gapped, while others are not - giving rise to the Fermi arc in ARPES measurements \cite{Damascelli03}. Cavity-mediated Amperean pairing in GaAs heterostructures shows the same phenomenology in a much simpler, cleaner system. By varying the electron density, it is possible to induce a change between a fully gapped low-density and a partially gapped high-density state.

\begin{figure}[t]
\centering
\includegraphics[width=0.4\textwidth]{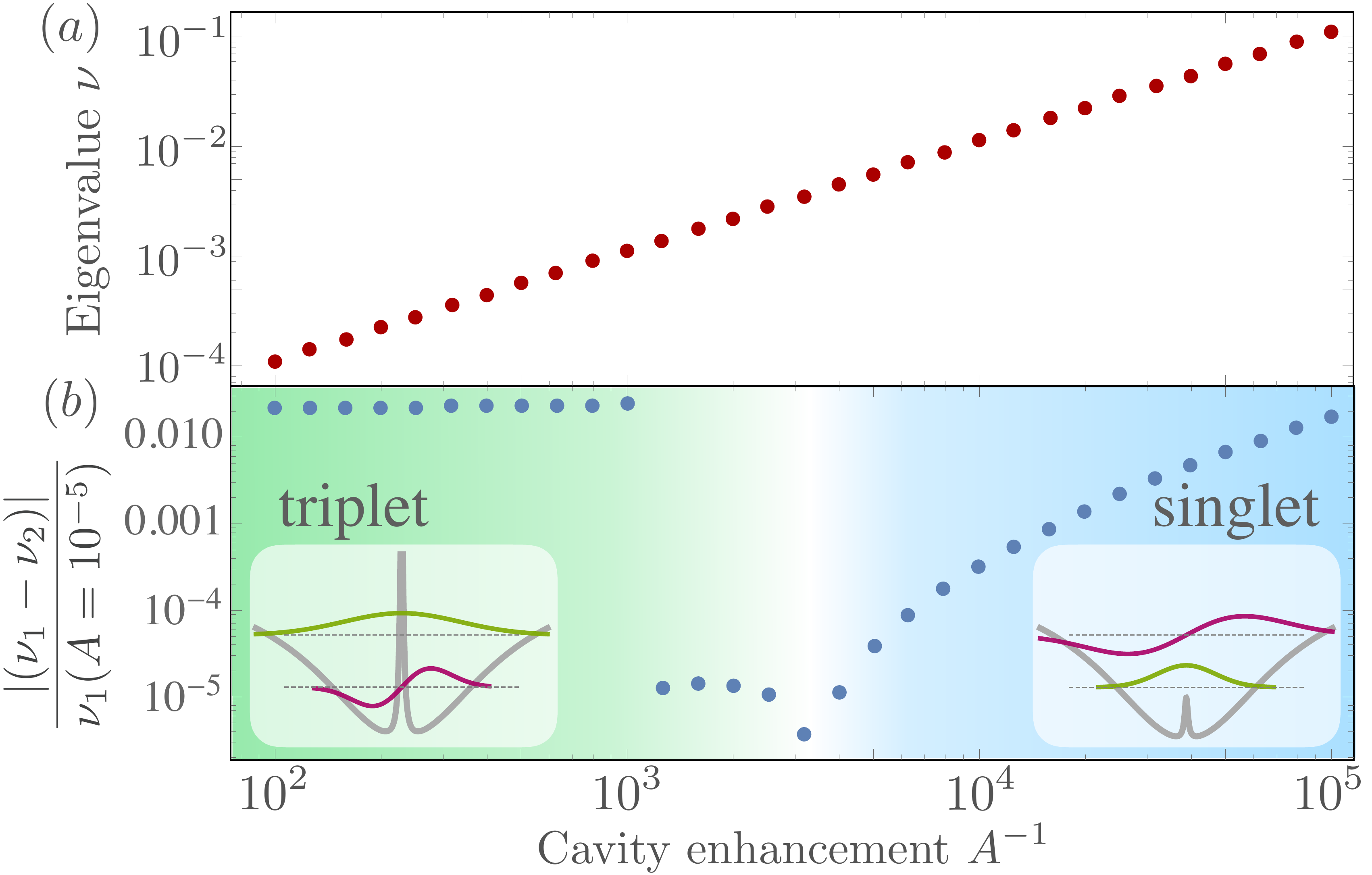}
\caption{
(a) The leading positive eigenvalue of Eq.~(\ref{eq.matrix}) is plotted as a function of the cavity enhancement $A^{-1}$, for $\mu = - 3.99 t$ ($i.e.$ $n_e \simeq 10^{12}$~cm$^{-1}$), and Coulomb repulsion appropriate for a 2D electron gas in GaAs (see SI).
(b) Absolute value of the difference between the leading $\nu_1$, and sub-leading eigenvalue $\nu_2$, normalized to the leading eigenvalue at $A = 10^{-5}$. At low enhancements, $\nu_1$ is negative (and the s-wave repulsive), resulting in large differences. The inset on the left sketches the effective interaction potential in this region in grey. The green wavefunction sketches the $s$-wave wavefunction which is strongly affected by the local repulsion, and thus suppressed energetically. As a consequence, the $p$-wave (red) is energetically favourable, and pairing occurs in the triplet channel. 
In the inset of the right, the repulsion is too weak to suppress the $s$-wave.
}
\label{fig.parameter-space}
\end{figure}

So far, we only considered the dominating eigenvalue obtained from Eq.~(\ref{eq.matrix}).
Yet the eigenvalue spectrum shown in Fig.~\ref{fig.setup}(c) shows a succession of eigenvalues with decreasing amplitude which are separated by less than 1 \% of their absolute values. The eigenfunctions corresponding to the two leading eigenvalues are shown in Fig.~\ref{fig.setup}(d), and correspond to a singlet order, with $\Delta^{(\vec{Q})}_{-\vec{p}} = \Delta^{(\vec{Q})}_{\vec{p}}$, and a triplet order with $\Delta^{(\vec{Q})}_{-\vec{p}} = - \Delta^{(\vec{Q})}_{\vec{p}}$, respectively. Since the cavity-mediated pairing potential of two electrons is long-range compared to electronic length scales, a node (as in triplet pairing) in the two-electron wavefunction only results in a small energy penalty, thus pushing the different orders towards degeneracy. Up to now, we neglected any electron-electron repulsion. Yet in a real material, the interaction potential between two electrons will look rather like the sketches inset in Fig.~\ref{fig.parameter-space}(b): the long-range cavity-mediated interaction dominates at large distances, but at short distances, other intrinsic interactions of the electron gas affect the pairing potential. The singlet wavefunction peaks at zero distance, and should be influenced strongly by local repulsive interactions. On the other hand, the triplet state vanishes at the origin, and should thus be affected less strongly. In the current situation, where different eigenvalues are separated by only few percents, these local interactions could decide the realised state. 
We will examine the simplest possible local interaction in the following: a constant Hubbard-$U$ describing a local repulsive electron interaction. We replace the original interaction in Eq.~(\ref{eq.matrix}) with $V^{(\vec{Q})}_{\vec{p}, \vec{p}'} + U$, and investigate the leading eigenvalues of the resulting effective interaction.
 
Fig.~\ref{fig.parameter-space}(a) shows the largest positive eigenvalue of the matrix eigenvalue equation~(\ref{eq.matrix}) as a function of the cavity enhancement $A^{-1}$, thus modulating the interaction strength.  
The eigenvalue grows linearly with $A^{-1}$, and shows no discernible structure. 
In Fig.~\ref{fig.parameter-space}(b), however, we show the normalised difference between the two dominating eigenvalues of the matrix eigenvalue equation~(\ref{eq.matrix}), revealing a structural change with $A^{-1}$. At very low enhancements, the difference is rather large. In this regime, the singlet state is negative (repulsive), and the system is dominated by triplet pairing. Then, at $A^{-1} \simeq 10^3$, the singlet state becomes attractive; the difference to the triplet state shrinks, and decreases until the cavity compression reaches $A^{-1} \simeq 2 \times 10^3$, where the singlet order becomes larger than the triplet component. Thus, with increasing coupling strength, the system switches from a low-temperature triplet regime to a singlet regime at strong couplings.
Using parameters for GaAs, however, this transition takes place at very low temperatures, and it appears difficult to observe the triplet phase in this system. It could possibly be observed in complex oxide interfaces \cite{Stemmer14} with smaller electron mobilities and larger electron densities $\sim 10^{15}$~cm$^{-2}$. Alternatively, it might be possible to suppress the singlet pairing by splitting the spin-dependent Fermi surfaces with a magnetic field.

Our results open a new route towards the optical manipulation of superconductivity.
The cavity-induced modification of the electronic ground state - similar to experiments in molecular systems \cite{Hutchinson12, Wang14, Orgiu15, Ebbesen16}, and related theoretical proposals to employ cavities to change molecular dynamics \cite{Feist15, Schachenmayer15, Galego15, Kowalewski16, Kowalewski16b, Herrera16, Galego16, Galego17} and ground states \cite{Flick17, Flick18} - could prove to be a very powerful tool to enhance or design coherent electronic states in two-dimensional materials. Furthermore, we have not considered possible external driving of the cavity. Due to optical nonlinearities of the electron gas, such driving would certainly influence the cavity-mediated interactions, and might possibly be able to stabilise electron pairing at higher temperatures. 

To conclude, we have investigated electron interactions mediated by a THz cavity strongly coupled to a two-dimensional electron system. These long-range interactions arise - very much like van der Waals or Casimir-Polder forces between macroscopic bodies \cite{Woods16} - from the exchange of virtual photons between spontaneous current fluctuations. 
Since the critical temperature scales approximately linearly with the cavity frequency, we suggest that Ref.~\cite{Zurich12}, for instance, conducted their measurements above $T_c$, and thus did not observe the effects we have predicted here.
We also did not investigate the effect of impurities, which could further reduce $T_c$.
Yet the counterrotating terms in the electron-photon coupling, which the present theory relies upon, were recently reported in \cite{Li18}, demonstrating the feasibility of observing cavity-mediated interactions.

\section{acknowledgements}
\begin{acknowledgements}
The authors would like to thank Giacomo Mazza for helpful discussions. 
The research leading to these results has received funding from the European Research Council under the European Union's Seventh Framework Programme (FP7/2007-2013) Grant Agreement No. 319286 Q-MAC. 
\end{acknowledgements}



\section*{Supplementary material}

# Supplementary Information: Cavity-mediated electron-photon superconductivity

Frank Schlawin[1], Andrea Cavalleri[1,2] and Dieter Jaksch[1]
[1] *Clarendon Laboratory, University of Oxford, Parks Road, Oxford OX1 3PU, United Kingdom** and*
[2] *Max Planck Institute for the Structure and Dynamics of Matter, Luruper Chaussee 149, 22761 Hamburg, Germany*

## I. THE MODEL

### A. Cavity field quantisation

The setup is sketched in Fig. 1(a) of the main text. We model the cavity as perfectly reflecting mirrors at $z = 0$ and $L$, with the 2D electron gas at the centre at $z_0 = L/2$. The tangential component of the electric field and the normal component of the magnetic fields must vanish at the mirrors' position. In addition, we impose periodic boundary conditions in the $xy$-plane. Given an electronic system filling the $xy$-plane with a spacing $a$ between the lattice sites, the size of the cavity in the plane is $L_x L_y = N a^2$. For the mean field, we shall consider the thermodynamic limit $N \to \infty$.

To derive expressions for the cavity field, we follow the derivation in [1]. The boundary conditions are satisfied by the vector potential of the form

$$\vec{A}' = \sum_{\vec{q}} \sqrt{\frac{\hbar}{\epsilon \epsilon_0 V \omega_q}} e^{i(q_x x + q_y y)} \begin{pmatrix} (b_x - b_x^\dagger) i \sin(\frac{\pi z}{L}) \\ (b_y - b_y^\dagger) i \sin(\frac{\pi z}{L}) \\ (b_z + b_z^\dagger) \cos(\frac{\pi z}{L}) \end{pmatrix}, \quad (S1)$$

where $b_n$ denotes the field amplitude in the $n$-th direction, $\epsilon$ the filling material dielectric constant, and $\epsilon_0$ the vacuum dielectric constant. Transversality imposes $\vec{q} \cdot \vec{A} = 0$, such that the $b_i$ are not linearly independent. Therefore, we write $\vec{q} = |q|(\sin\theta\cos\phi, \sin\theta\sin\phi, \cos\theta)$, and rotate to a new basis, in which the $z$-component is parallel to $\vec{q}$, with the rotation matrix

$$O^q = \begin{pmatrix} \cos\theta\cos\phi & \cos\theta\sin\phi & -\sin\theta \\ -\sin\phi & \cos\phi & 0 \\ \sin\theta\cos\phi & \sin\theta\sin\phi & \cos\theta \end{pmatrix}. \quad (S2)$$

The new eigenmodes are then given by $\vec{u}_{\vec{q}s} = \sum_i \hat{e}_i O^q_{si} u_{\vec{q}i}$, and we obtain the two transverse field modes

$$\vec{A} = i \sum_{\vec{q},s} \sqrt{\frac{\hbar}{\epsilon \epsilon_0 V \omega_q}} \hat{e}_{\vec{q},s} e^{i(q_x x + q_y y)} \left( a_{\vec{q},s} + a^\dagger_{-\vec{q},s} \right), \quad (S3)$$

with the polarisation vector

$$\hat{e}_{\vec{q},1} = \frac{\vec{q}}{|\vec{q}|}, \quad (S4)$$

which is called the transverse electric mode. Here, we have set $\pi z_0 / L = \pi/2$, i.e. the electron system is placed at the maximum of the cavity field.

The second polarisation is given by

$$\hat{e}_{\vec{q},2} = \hat{e}_{\vec{q},1} \times \hat{e}_z \quad (S5)$$

and is called the transverse magnetic mode. The $z$-component of the field can be neglected, and we obtain the vector potential

$$\vec{A}(\vec{r}) = i \sum_{\vec{q},s} \sqrt{\frac{\hbar}{\epsilon \epsilon_0 V \omega_q}} e^{i(q_x x + q_y y)} \hat{e}_{\vec{q},s} \left( a_{\vec{q},s} + a^\dagger_{-\vec{q},s} \right). \quad (S6)$$

Note that the transverse magnetic mode also has a gapless mode with $k_z = 0$, which is however polarised along the $z$-direction [2], and does not couple to the electronic system in the present setup.

### B. Coupling to electron field

The paramagnetic Hamiltonian couples the cavity field to the current operator of the electron system, and reads in real space

$$H_{int} = \sum_{\vec{n}} \vec{j}(\vec{r}_{\vec{n}}) \cdot \vec{A}((\vec{r}_{\vec{n}+1} - \vec{r}_{\vec{n}})/2), \quad (S7)$$

where $\vec{n}$ denotes the summation over the lattice sites. In a rectangular lattice with nearest-neighbour hopping, the current operator is given by

$$\vec{j}_n = \frac{iaet}{\hbar} \sum_{i=x,y} \sum_{\vec{n},\sigma} \hat{e}_i \left( c^\dagger_{\vec{n}+1_i,\sigma} c_{\vec{n},\sigma} - c^\dagger_{\vec{n},\sigma} c_{\vec{n}+1_i,\sigma} \right), \quad (S8)$$

where $\vec{n} + 1_x = (n_x + 1, n_y)$ and $\vec{n} + 1_y = (n_x, n_y + 1)$. Using the form of the vector potential (S6), the interaction Hamiltonian in $k$-space then reads

$$H_{int} = \sum_{\vec{k},\sigma,\vec{q},s} \frac{g^{(\vec{q})}_{\vec{k},s}}{\sqrt{N}} \left( a_{\vec{q},s} + a^\dagger_{-\vec{q},s} \right) c^\dagger_{\vec{k}+\vec{q},\sigma} c_{\vec{k},\sigma}, \quad (S9)$$

with

$$g^{(\vec{q})}_{\vec{k},1} = i\tilde{g}_0 \frac{\hat{e}_{\vec{q},1,x} \sin[(k_x + q_x/2)] + \hat{e}_{\vec{q},1,y} \sin[(k_y + q_y/2)]}{\left(1 + \left(\frac{c\vec{q}}{\omega_0}\right)^2\right)^{1/4}}, \quad (S10)$$

and

$$g^{(\vec{q})}_{\vec{k},2} = i\tilde{g}_0 \frac{\hat{e}_{\vec{q},2,x} \sin[(k_x + q_x/2)] + \hat{e}_{\vec{q},2,y} \sin[(k_y + q_y/2)]}{\left(1 + \left(\frac{c\vec{q}}{\omega_0}\right)^2\right)^{1/4}}, \quad (S11)$$

---

* frank.schlawin@physics.ox.ac.uk

where in our setup

$$\tilde{g}_0 = t\sqrt{\frac{4\alpha}{\sqrt{\epsilon}}}, \tag{S12}$$

$$\simeq 0.1t. \tag{S13}$$

Here, $\alpha \simeq 1/137$ denotes the fine structure constant, and we used $\epsilon = 13$ for GaAs (see table I). This coefficient is similar to the one obtained in [3] for the coupling of cyclotron transitions to a cavity field. The coefficients obey the symmetry $g^{(-\vec{q})}_{-\vec{k},s} = g^{(\vec{q})}_{\vec{k},s}$ and $g^{(-\vec{q})}_{\vec{k},s} \simeq g^{(\vec{q})*}_{\vec{k},s}$.

This is the cavity coupling for a half-wavelength cavity with $\omega_0 = c\pi/L_z$. In the case of a nanoplasmonic cavity, the field is further enhanced below the free-space limit. For instance, in Ref. [4], the cavity volume is estimated to be $V = 2.5 \times 10^{-5} \times (\lambda_0/2\sqrt{\epsilon})^3$, where $\lambda_0$ is the vacuum cavity wavelength. It is this further compression of the cavity field that enables the experiments to reach the ultrastrong coupling regime, where the coupling becomes comparable to the cavity frequency. To describe this situation phenomenologically (*i.e.* without ab initio simulation of the cavity field), we rescale $\tilde{g}_0$ by the reduction of the mode volume below the $\lambda^3$-limit,

$$g_0 = \frac{\tilde{g}_0}{\sqrt{A}}, \tag{S14}$$

where we define the reduction of the cavity below the far field limit $A$.

## II. PARAMETER VALUES FOR GAAS HETEROSTRUCTURES

| | |
|---|---|
| Electron band mass | $m^* = 0.07\, m_e$ |
| Lattice constant | $a = 5.6$ Å |
| Dielectric constant | $\epsilon = 13$ |
| Electron density | $n_e = 3.6 \times 10^{11}$ cm$^{-2}$ |
| Cavity frequency | $\omega_0 = 2\pi \times 5$ THz |

TABLE I. Parameters of GaAs used for the estimation of the critical temperature.

The parameters are obtained from reported values in [4, 7–9], and are given in table I. From these, we calculate the transfer integral

$$t = \frac{\hbar^2}{2m^*a^2}, \tag{S15}$$

and the Fermi wavevector (which we measure in units of the lattice constant $a$)

$$k_f a = \sqrt{2\pi n_e}\, a \simeq 0.084. \tag{S16}$$

For low filling ratios, the electron dispersion is approximately quadratic,

$$\epsilon_k \approx t a^2 \left(k_x^2 + k_y^2\right) - \tilde{\mu} \tag{S17}$$

$$= \frac{\hbar^2}{2m^*}\left(k_x^2 + k_y^2\right) - \tilde{\mu}, \tag{S18}$$

with $\tilde{\mu} = 4t + \mu$. We use Eqs. (S16) and (S18) to relate the electron density to the chemical potential,

$$n_e = \frac{m^*}{\pi\hbar^2}(4t + \mu). \tag{S19}$$

The speed of light in GaAs in given by

$$c = c_0/\sqrt{\epsilon}, \tag{S20}$$

where $c_0 = 3 \times 10^8$ m/s is the vacuum speed of light.

In the case of a two-dimensional electron gas GaAs heterostructures, the screened Coulomb repulsion is given by [9]

$$V_C(\vec{k}) = \frac{1}{2\epsilon\epsilon_0 N a^2}\frac{e^2}{|\vec{k}| + k_{TF}}, \tag{S21}$$

with the vacuum permittivity $\epsilon_0$ and the inverse screening length $k_{TF} = me^2/(2\pi\epsilon\epsilon_0\hbar^2)$, we obtain an approximate Hubbard-U (for $\vec{k} = 0$), $U \simeq 0.25t$.

## III. WOLF-SCHRIEFFER TRANSFORMATION

To obtain the effective cavity-mediated interaction, we eliminate the interaction Hamiltonian (S9) by a Wolf-Schrieffer transformation, *i.e.* we rotate to a new frame with Hamiltonian

$$H' \equiv e^S \left(H_0 + H_f + H_{int}\right) e^{-S} \tag{S22}$$

$$= H_0 + H_f + H_{int} + \left[S, H_0 + H_f + H_{int}\right]$$

$$+ \frac{1}{2}\left[S, \left[S, H_0 + H_f\right]\right] + \ldots \tag{S23}$$

where we choose $S$, such that

$$\left[S, H_0 + H_f\right] = -H_{int}, \tag{S24}$$

and thereby eliminate the coupling to the cavity to leading order. This is the case for

$$S = \sum_{\vec{k},\vec{q},\sigma,s} \frac{g^{(\vec{q})}_{\vec{k},s}}{\sqrt{N}}\left(x^{(s)}_{\vec{k},\vec{q}} a_{\vec{q},s} + y^{(s)}_{\vec{k},\vec{q}} a^\dagger_{-\vec{q},s}\right) c^\dagger_{\vec{k}+\vec{q},\sigma} c_{\vec{k},\sigma}, \tag{S25}$$

with

$$x^{(s)}_{\vec{k},\vec{q}} = -\frac{1}{\epsilon_{\vec{k}+\vec{q}} - \epsilon_{\vec{k}} - \hbar\omega_{\vec{q}}} \tag{S26}$$

and

$$y^{(s)}_{\vec{k},\vec{q}} = -\frac{1}{\epsilon_{\vec{k}+\vec{q}} - \epsilon_{\vec{k}} + \hbar\omega_{\vec{q}}}. \tag{S27}$$

From Eq. (S23), we then obtain

$$H_{\text{eff}} = \frac{1}{2N}\sum_{\vec{k},\vec{k}',\vec{q}}\sum_{\sigma,\sigma',s}\frac{2\hbar\omega_q}{(\epsilon_{\vec{k}+\vec{q}}-\epsilon_{\vec{k}})^2 - (\hbar\omega_q)^2} g^{(\vec{q})}_{\vec{k},s} g^{(-\vec{q})}_{\vec{k}',s} c^\dagger_{\vec{k}+\vec{q},\sigma} c_{\vec{k},\sigma} c^\dagger_{\vec{k}'-\vec{q},\sigma'} c_{\vec{k}',\sigma'}. \tag{S28}$$

Keeping in mind that $\hbar\omega_q \gg |\epsilon_{\vec{k}} - \epsilon_{\vec{k}'}|$, we can always replace

$$\frac{2\hbar\omega_q}{(\epsilon_{\vec{k}+\vec{q}}-\epsilon_{\vec{k}})^2 - (\hbar\omega_q)^2} \simeq -\frac{2}{\hbar\omega_{\vec{q}}}. \tag{S29}$$

## A. Change of electron dispersion

In tracing out the photon field, one also obtains a renormalisation of the electronic band structure due to scattering processes, when $\vec{k}' = \vec{k} + \vec{q}$ and $\sigma' = \sigma$. We obtain

$$H_{\text{St}} = \sum_{\vec{k},\sigma} h_{\vec{k}} c^\dagger_{\vec{k}\sigma} c_{\vec{k}\sigma}, \tag{S30}$$

with

$$h_{\vec{k}} = -\frac{1}{N} \sum_{\vec{q},s} \frac{g^{(\vec{q})}_{\vec{k}-\vec{q},s} g^{(-\vec{q})}_{\vec{k},s}}{\hbar \omega_q} \tag{S31}$$

$$= -\frac{g_0^2}{N\hbar\omega_0} \left(\frac{\omega_0}{c}\right)^2 \sum_{\vec{q}} \frac{\sum_{i=x,y} \sin^2[(k_i a - q_i a/2)]}{(\omega_0/c)^2 + q^2}. \tag{S32}$$

It can be estimated to be on the order of

$$h_{\vec{k}} \simeq -\pi V_0 k_f^2 \left(\frac{a\omega_0}{c}\right)^2 \int_0^{q_{\max}} dq \frac{q}{(\omega_0/c)^2 + q^2} \tag{S33}$$

$$\simeq -\pi V_0 k_f^2 \left(\frac{a\omega_0}{c}\right)^2 \log\left[1 + \left(\frac{c}{\omega_0} q_{\max}\right)\right], \tag{S34}$$

where $q_{\max}$ denotes a cutoff frequency for the in-plane photon momenta supported by the cavity (see also section IV D). In the first line, we approximated the numerator of Eq. (S32) by $\sim 2k_f^2 a^2$. The additional factor $2\pi$ stems from the integration over the azimuthal angle. Using values appropriate for GaAs, we obtain $h_{\vec{k}} \sim 1.25 \times 10^{-8} A^{-1}$. Thus, even if the mode volume compression is as small as $A = 10^{-5}$, the change in the electron dispersion would on the order of 0.1 %. It can therefore be neglected.

## B. The diamagnetic interaction

As shown theoretically in [3], and observed experimentally in [7], the diamagnetic interaction, which can be neglected in most cases of interest in the optical regime, can become relevant or even dominant at strong resonant coupling. The diamagnetic interaction Hamiltonian reads

$$H_{\text{dia}} = \sum_{\vec{k},\vec{q},\vec{q}',s,\sigma} \gamma_{\vec{k},\vec{q},\vec{q}'} \left(a_{\vec{q}'-\vec{q},s} + a^\dagger_{\vec{q}-\vec{q}',s}\right)\left(a_{\vec{q},s} + a^\dagger_{-\vec{q},s}\right) c^\dagger_{\vec{k}+\vec{q},\sigma} c_{\vec{k}\sigma}, \tag{S35}$$

where

$$\gamma_{\vec{k},\vec{q},\vec{q}'} = -\frac{t}{A} \left(\frac{ea}{\hbar}\right)^2 \sqrt{\frac{\hbar}{\epsilon V \omega_q}} \sqrt{\frac{\hbar}{\epsilon V \omega_{q'}}}. \tag{S36}$$

The prefactor can be estimated as (see section IV D)

$$\left|\gamma_{\vec{k},\vec{q},\vec{q}'}\right| \sim \frac{t}{A} \frac{4\alpha}{\sqrt{\epsilon}} \left(\frac{a\omega_0}{c}\right)^2 \log\left[1 + \left(\frac{c}{\omega_0} q_{\max}\right)\right] \sim 10^{-8} \frac{t}{A}, \tag{S37}$$

and can only give rise to minimal changes of the electron self-energy. Again, it can be safely neglected, as long as the cavity is in its ground state.

## IV. MEAN FIELD THEORY

### A. The pair density wave

Considering only the attractive Amperean channel in the nesting wavevectors $\vec{Q} = k_0(\pm 1, \pm 1)$, the electronic Hamiltonian reads

$$H = \sum_{\vec{k},\sigma} \epsilon_{\vec{k}} c^\dagger_{\vec{k},\sigma} c_{\vec{k},\sigma} + \frac{1}{2N} \sum_{\vec{Q},\vec{p},\vec{p}'} \sum_{\sigma,\sigma'} V^{(\vec{Q})}_{\vec{p},\vec{p}'} c^\dagger_{\vec{Q}+\vec{p},\sigma} c_{\vec{Q}+\vec{p}',\sigma} c^\dagger_{\vec{Q}-\vec{p},\sigma'} c_{\vec{Q}-\vec{p}',\sigma'}. \tag{S38}$$

We define the order parameters

$$\Delta^{(\vec{Q})}_{\sigma,\sigma'}(\vec{p}) = \frac{1}{N} \sum_{\vec{p}'} V^{(\vec{Q})}_{\vec{p},\vec{p}'} \left\langle c_{\vec{Q}-\vec{p}',\sigma'} c_{\vec{Q}+\vec{p}',\sigma} \right\rangle, \tag{S39}$$

and arrive at the mean field Hamiltonian

$$H_{\text{MF}} = \frac{1}{2} \sum_{\vec{Q},\vec{p}} \begin{pmatrix} c^\dagger_{\vec{Q}+\vec{p},\uparrow} & c^\dagger_{\vec{Q}+\vec{p},\downarrow} & c_{\vec{Q}-\vec{p},\uparrow} & c_{\vec{Q}-\vec{p},\downarrow} \end{pmatrix} \begin{pmatrix} \epsilon_{\vec{Q}+\vec{p}} & 0 & \Delta^{(\vec{Q})}_{\uparrow\uparrow}(\vec{p}) & \Delta^{(\vec{Q})}_{\uparrow\downarrow}(\vec{p}) \\ 0 & \epsilon_{\vec{Q}+\vec{p}} & \Delta^{(\vec{Q})}_{\downarrow\uparrow}(\vec{p}) & \Delta^{(\vec{Q})}_{\downarrow\downarrow}(\vec{p}) \\ \Delta^{(\vec{Q})*}_{\uparrow\uparrow}(\vec{p}) & \Delta^{(\vec{Q})*}_{\downarrow\uparrow}(\vec{p}) & -\epsilon_{\vec{Q}-\vec{p}} & 0 \\ \Delta^{(\vec{Q})*}_{\uparrow\downarrow}(\vec{p}) & \Delta^{(\vec{Q})*}_{\downarrow\downarrow}(\vec{p}) & 0 & -\epsilon_{\vec{Q}-\vec{p}} \end{pmatrix} \begin{pmatrix} c_{\vec{Q}+\vec{p},\uparrow} \\ c_{\vec{Q}+\vec{p},\downarrow} \\ c^\dagger_{\vec{Q}-\vec{p},\uparrow} \\ c^\dagger_{\vec{Q}-\vec{p},\downarrow} \end{pmatrix} + \sum_{\vec{Q},\vec{p}} \epsilon_{\vec{Q}-\vec{p}}. \tag{S40}$$

This representation contains singlet and triplet components. The dominating symmetry cannot be deduced in this form.

### B. Single mean field

We consider only one order parameter, in which case the mean field Hamiltonian reduces to

$$H'_{\text{MF}} = \sum_{\vec{Q},\vec{p}} \begin{pmatrix} c^\dagger_{\vec{Q}+\vec{p},\sigma} & c_{\vec{Q}-\vec{p},\sigma'} \end{pmatrix} \begin{pmatrix} \epsilon_{\vec{Q}+\vec{p}} & \Delta^{(\vec{Q})}_{\sigma\sigma'}(\vec{p}) \\ \Delta^{(\vec{Q})*}_{\sigma\sigma'}(\vec{p}) & -\epsilon_{\vec{Q}-\vec{p}} \end{pmatrix} \begin{pmatrix} c_{\vec{Q}+\vec{p},\sigma} \\ c^\dagger_{\vec{Q}-\vec{p},\sigma'} \end{pmatrix}. \tag{S41}$$

It can be diagonalised straightforwardly, and we obtain

$$H'_{\rm MF} = \sum_{\vec{Q},\vec{p}} \begin{pmatrix} \gamma^\dagger_{\vec{Q}+\vec{p},\sigma} & \gamma_{\vec{Q}-\vec{p},\sigma'} \end{pmatrix} \begin{pmatrix} E_+ & 0 \\ 0 & E_- \end{pmatrix} \begin{pmatrix} \gamma_{\vec{Q}+\vec{p},\sigma} \\ \gamma^\dagger_{\vec{Q}-\vec{p},\sigma'} \end{pmatrix}, \quad (S42)$$

with $E_\pm = \frac{1}{2}(\epsilon_{\vec{Q}+\vec{p}} - \epsilon_{\vec{Q}-\vec{p}}) \pm \sqrt{(\epsilon_{\vec{Q}+\vec{p}} + \epsilon_{\vec{Q}-\vec{p}})^2/4 + |\Delta^{(\vec{Q})}(\vec{p})|^2}$. Note that in contrast to regular BCS theory, this Hamiltonian has no particle-hole symmetry, $E_- \neq -E_+$.

The gap equation reads

$$\Delta^{(\vec{Q})}_{\sigma,\sigma'}(\vec{p}) = -\frac{1}{N} \sum_{\vec{p}'} V^{(\vec{Q})}_{\vec{p},\vec{p}'} \left\langle c_{\vec{Q}-\vec{p}',\sigma'} c_{\vec{Q}+\vec{p}',\sigma} \right\rangle \quad (S43)$$

$$= -\frac{1}{N} \sum_{\vec{p}'} V^{(\vec{Q})}_{\vec{p},\vec{p}'} \frac{\Delta(\vec{p}')}{2\sqrt{\zeta^2_{\vec{p}'} + |\Delta(\vec{p}')|^2}} \left( f_{\vec{p}'} + f_{-\vec{p}'} - 1 \right), \quad (S44)$$

where $f_{\vec{p}} = (\exp(E_+(\vec{p})/k_BT)+1)^{-1}$, and $\zeta_p = (\epsilon_{\vec{Q}+\vec{p}} + \epsilon_{\vec{Q}-\vec{p}})/2$. Due to the lack of particle-hole symmetry, the expression does not simplify to the usual gap equation, since

$$f_{-\vec{p}} = \frac{1}{e^{E_+(-\vec{p})/kBT}} = \frac{1}{e^{(E_+(\vec{p})-(\epsilon_{\vec{Q}+\vec{p}}-\epsilon_{\vec{Q}-\vec{p}}))/kBT}} \neq f_{\vec{p}}. \quad (S45)$$

Thus, in the continuous limit we obtain

$$\Delta^{(\vec{Q})}_{\sigma,\sigma'}(\vec{p}) = -\frac{|V_0|}{2(2\pi)^2} \int d^2p' \frac{1}{2} \frac{\cos(p_x+p'_x)a + \cos(p_y+p'_y)a - \cos 2Q_xa - \cos 2Q_ya}{(\omega_0/c)^2 + |\vec{p}'-\vec{p}|^2} \frac{\Delta^{(\vec{Q})}_{\sigma,\sigma'}(\vec{p}')}{\sqrt{\zeta^2_{\vec{p}'} + |\Delta_{\vec{p}'}|^2}} \left( f_{\vec{p}'} + f_{-\vec{p}'} - 1 \right). \quad (S46)$$

As we will describe in the next section, the deviation from a conventional BCS theory are negligibly small; at least, when we are determining the critical temperature.

### C. Critical temperature

We wish to switch the integration over the full Brillouin zone in Eq. (S46) into an integration along the Fermi surface, and one integration along the energy axis $\zeta$. To find deviations from conventional BCS theory, we write

$$f_{\vec{p}'} + f_{-\vec{p}'} - 1 = \frac{1}{e^{(\zeta+\delta)/k_BT}+1} + \frac{1}{e^{(\zeta-\delta)/k_BT}+1} - 1, \quad (S47)$$

where $\delta$ denotes the deviation of the quasiparticle dispersion from $\zeta$. Expanding around $\delta = 0$ (at the Fermi surface), we find

$$f_{\vec{p}'} + f_{-\vec{p}'} - 1 \simeq -\tanh\left(\frac{\zeta}{2k_BT}\right) + 2\left(\frac{\delta}{k_BT}\right)^2 \frac{\sinh^4(\zeta/(2k_BT))}{\sinh^3(\zeta/(k_BT))} + \ldots \quad (S48)$$

Linearising Eq. (S46) near $T_c$, and switching the integration variables to an integration along the energy axis, and along the Fermi surface, we obtain the eigenvalue equation (6) of the main text. The energy integration yields for the first term in Eq. (S48) the well-known result

$$\int_0^{\hbar\omega_c} \frac{d\zeta}{\zeta} \tanh\left(\frac{\zeta}{2k_BT_c}\right) \simeq \ln\left(\frac{1.1.3\hbar\omega_c}{k_BT_c}\right), \quad (S49)$$

and we find that corrections to this term are suppressed exponentially,

$$\int_0^{\hbar\omega_c} \frac{d\zeta}{\zeta} \frac{\sinh^4(\zeta/(2k_BT))}{\sinh^3(\zeta/(k_BT))} \sim \exp\left(-\frac{\hbar\omega_c}{k_BT_c}\right). \quad (S50)$$

Hence, for the determination of the critical temperature, we can neglect the deviation from particle-hole symmetry, and evaluate Eq. (S46) with $f_{\vec{p}'} + f_{-\vec{p}'} - 1 \simeq -\tanh(\zeta/(2k_BT))$. The critical temperature is then connected to the largest eigenvalue of the linearised gap equation by [5]

$$k_BT_c = 1.13\hbar\omega_c e^{-1/\nu}, \quad (S51)$$

where $\nu$ is a solution of the eigenvalue equation (6) of the main text. Using $A = 5\times 10^{-5}$ and $\omega_c = \omega_0$, we obtain $T_c = 2$ K. For $A = 10^{-5}$, this would even increase to $T_c = 68$ K (However, in this case the eigenvalues grow to order of unity, $\nu \sim 1$, and our weak-coupling approach may no longer be adequate). This approximation cannot account for a finite range of photon momenta in the cavity. So we compare these results in the following with an alternative calculation in section IV D.

### D. Check: long-range approximation

Here, we consider the full gap equation for $s$-wave pairing symmetry. The $p$-wave case could be treated analogously, but here we are only interested in an estimate for the critical temperature.

At finite temperature, the gap equation is given by

$$\Delta^{(\vec{Q})}(\vec{p}) = \sum_{\vec{p}'} \frac{V^{(\vec{Q})}_{\vec{p},\vec{p}'}}{2N} \frac{\Delta^{(\vec{Q})}(\vec{p}')}{\sqrt{\zeta^2_{\vec{p}'} + |\Delta^{(\vec{Q})}(\vec{p}')|^2}} \left(f_{\vec{p}'} + f_{-\vec{p}'} - 1\right). \tag{S52}$$

In the thermodynamic limit, We obtain Eq. (S46). To solve this integral equation analytically, we note that $c/(a\omega_0) \sim 10^4$ in the THz regime, so the interaction $V_{\vec{p},\vec{p}'}$ decays very quickly in $k$-space (*i.e.* it is long-range in real space). Thus, only a very short range of wavevectors will be affected, and we can write approximately,

$$\frac{1}{(\omega_0/c)^2 + |\vec{p}' - \vec{p}|^2} \simeq a_0 \delta(\vec{p}' - \vec{p}), \tag{S53}$$

where $a_0$ is chosen such that the both sides of Eq. (S53) integrated over the first Brillouin zone yield the same result, *i.e.*

$$a_0 = \int_{BZ} d^2q \, \frac{1}{(\omega_0/c)^2 + |\vec{q}|^2} \tag{S54}$$

$$\simeq 2\pi \int_0^{q_{\max}} dq \, \frac{q}{(\omega_0/c)^2 + q^2}, \tag{S55}$$

where we introduced a momentum cutoff $q_{\max}$ for the in-plane photon momenta supported by the cavity in the second line which we discuss below. We then obtain

$$a_0 = \pi \log\left[1 + \left(\frac{c}{\omega_0} q_{\max}\right)^2\right]. \tag{S56}$$

The factor $2\pi$ in Eq. (S55) stems from the integration over the azimuthal angle. The momentum cutoff $q_{\max}$ evidently depends on the cavity geometry. Our present model of two infinite cavity walls sustains arbitrary $q$-values, and we can set $q_{\max} = \pi/a$. In this case, we obtain $\log(1 + (\pi c/(a\omega_0))^2) \simeq 20$ (for the values in GaAs). In practice, however, only a finite range of photon momenta will be sustained. To estimate a lower bound for these, we note that the *minimal* photon momentum in a cavity of length $L$ in the $x/y$-plane should be $q_{\min} = \pi/L$. In [4], this length is $L = 45$ $\mu$m, yielding $q_{\min}a = 3.3 \times 10^{-5}$, and thus $\log[\ldots] \geq \log(1 + (1/3)^2) = 0.1$. In [6], the cavity length is given by $L = 2$ $\mu$m, thus $q_{\min} = 8.7 \times 10^{-4}$, and $\log[\ldots] \geq 2.9$. In the following, we assume the logarithm to be on the order of unity,

$$a_0 \simeq \pi \tag{S57}$$

The approximation (S53) turns Eq. (S46) into a simple inversion problem, and we obtain at zero temperature

$$\Delta^{(\vec{Q})}_0(\vec{p}) = \sqrt{\left(\frac{\pi V_0 \sum_{i=x,y} (\cos 2p_i a - \cos 2Q_i a)}{4(2\pi)^2}\right)^2 - \zeta^2_{\vec{p}}}, \tag{S58}$$

for $|\zeta_{\vec{p}}| < \pi V_0 (\cos 2p_i a - \cos 2Q_i a)/(4(2\pi)^2)$, and zero otherwise, and the critical temperature

$$k_B T_c \simeq \frac{\Delta^{(\vec{Q})}_0}{2}, \tag{S59}$$

where $\Delta^{(\vec{Q})}_0$ is given by Eq. (S58) evaluated at the Fermi surface. For typical values in GaAs heterostructures and using a cavity compression $V \sim \lambda^3/(2 \times 10^4)$, we obtain $V_0 \sim 10^{-3}t$, and the critical temperature

$$T_c \simeq 2 \times 10^{-4} \frac{K}{A}. \tag{S60}$$

Thus, the cavity-mediated interaction can give rise to superconducting condensation in the milliKelvin regime for a wavelength-limited cavity. This temperature is already orders of magnitude larger than those predicted for the free space interaction. The additional enhancement of the vacuum field in nanocavities then pushes the critical temperature into an experimentally accessible regime. For $A = 2 \times 10^{-4}$, we thus obtain $T_c \sim 1$ K, and for $A = 10^{-5}$, $T_c \sim 20$ K. This is roughly in agreement with the previous estimation in section IV C.

## V. QUASIPARTICLE HAMILTONIAN

To determine the energy dispersion, we diagonalize the Hamiltonian [10]

$$H_{\mathrm{MF}} = \begin{pmatrix} \epsilon_{\vec{k}} & \Delta^{(\vec{Q}_1)}_{\vec{k}} & \Delta^{(-\vec{Q}_1)}_{\vec{k}} & \Delta^{(\vec{Q}_2)}_{\vec{k}} & \Delta^{(-\vec{Q}_2)}_{\vec{k}} \\ \Delta^{(\vec{Q}_1)*}_{\vec{k}} & -\epsilon_{2\vec{Q}_1-\vec{k}} & 0 & 0 & 0 \\ \Delta^{(-\vec{Q}_1)*}_{\vec{k}} & 0 & -\epsilon_{-2\vec{Q}_1-\vec{k}} & 0 & 0 \\ \Delta^{(\vec{Q}_2)*}_{\vec{k}} & 0 & 0 & -\epsilon_{2\vec{Q}_2-\vec{k}} & 0 \\ \Delta^{(-\vec{Q}_2)*}_{\vec{k}} & 0 & 0 & 0 & -\epsilon_{-2\vec{Q}_2-\vec{k}} \end{pmatrix}, \tag{S61}$$

where we define $\vec{Q}_1 = k_0(1, 1)$ and $\vec{Q}_2 = k_0(1, -1)$. The length $k_0$ is obtained from the requirement $\epsilon_{\vec{Q}} = 0$. This Hamiltonian contains the mean field values $\Delta^{(\vec{Q})}_{\vec{k}}$ at all four nesting vectors. We approximate the mean fields by exponentials

$$\Delta^{(\vec{Q})}_{\vec{k}} = \Delta_0 \exp(-(\vec{k} - \vec{Q})^2/(2\sigma^2)), \tag{S62}$$

where the width $\sigma$ is estimated from a fit to the leading eigenfunction [see Fig. 1(d) of the main text] at the given electron density, but the results do not depend sensitively on this width. Furthermore, we fix $\Delta_0 = 10^{-3}t$ (corresponding to a critical temperature in the low-Kelvin regime).




\begin{thebibliography}{99}

\bibitem{Carbotte90} J. P. Carbotte, \href{https://link.aps.org/doi/10.1103/RevModPhys.62.1027}{Rev. Mod. Phys. \textbf{62}, 1027 (1990).}

\bibitem{Lee06} P. A. Lee, N. Nagaosa and X.G. Wen, \href{https://link.aps.org/doi/10.1103/RevModPhys.78.17}{Rev. Mod. Phys. \textbf{78}, 17 (2006).}

\bibitem{Holstein73} T. Holstein, R. E. Norton and P. Pincus, \href{https://link.aps.org/doi/10.1103/PhysRevB.8.2649}{Phys. Rev. B \textbf{8}, 2649 (1973).}

\bibitem{Reizer89a} M. Yu. Reizer, \href{https://link.aps.org/doi/10.1103/PhysRevB.39.1602}{Phys. Rev. B \textbf{39}, 1602 (1989).}

\bibitem{Reizer89b}M. Yu. Reizer, \href{https://link.aps.org/doi/10.1103/PhysRevB.40.11571}{Phys. Rev. B \textbf{40}, 11571 (1989).}

\bibitem{Khveshchenko93a} D. V. Kveshchenko, \href{https://link.aps.org/doi/10.1103/PhysRevB.47.3446}{Phys. Rev. B \textbf{47}, 3446 (1993).}

\bibitem{Khveshchenko93} D. V. Kveshchenko, R. Hlubina and T. M. Rice, \href{https://link.aps.org/doi/10.1103/PhysRevB.48.10766}{Phys. Rev. B \textbf{48}, 10766--10776 (1993).}

\bibitem{Lee07} S.-S. Lee, P. A. Lee and T. Senthil, \href{https://link.aps.org/doi/10.1103/PhysRevLett.98.067006}{Phys. Rev. Lett. \textbf{98}, 067006 (2007).}

\bibitem{Galitski08} V. Galitski and Y. B. Kim, \href{https://link.aps.org/doi/10.1103/PhysRevLett.99.266403}{Phys. Rev. Lett. \textbf{99}, 266403 (2007).}

\bibitem{Lee14} P. A. Lee, \href{https://link.aps.org/doi/10.1103/PhysRevX.4.031017}{Phys. Rev. X \textbf{4}, 031017 (2014).}

\bibitem{Kargarian16} M. Kargarian, D. K. Efimkin and V. Galitski, \href{https://link.aps.org/doi/10.1103/PhysRevLett.117.076806}{Phys. Rev. Lett. \textbf{117}, 076806 (2016).}

\bibitem{Zurich12} G. Scalari et al., \href{http://science.sciencemag.org/content/335/6074/1323}{Science \textbf{335}, 1323 (2012).}

\bibitem{Liu14} X. Liu et al., \href{http://dx.doi.org/10.1038/nphoton.2014.304}{Nature Photon. \textbf{9}, 30--34 (2015).}

\bibitem{Smolka14} S. Smolka et al., \href{http://science.sciencemag.org/content/346/6207/332}{Science \textbf{346}, 332--335 (2014).}

\bibitem{Maissen14} C. Maissen et al., \href{https://link.aps.org/doi/10.1103/PhysRevB.90.205309}{Phys. Rev. B \textbf{90}, 205309 (2014).}

\bibitem{Zhang16} Q. Zhang et al., \href{http://dx.doi.org/10.1038/nphys3850}{Nature Phys. \textbf{12}, 1005-1011 (2016).}

\bibitem{Bayer17} A. Bayer et al., \href{https://doi.org/10.1021/acs.nanolett.7b03103}{Nano Lett. \textbf{17}, 6340 (2017).}

\bibitem{Keller17} J. Keller et al., \href{https://doi.org/10.1021/acs.nanolett.7b03228}{Nano Letters \textbf{17}, 7410 (2017).}

\bibitem{Laussy10} F. P. Laussy, A. V. Kavokin and I. A. Shelykh, \href{https://link.aps.org/doi/10.1103/PhysRevLett.104.106402}{Phys. Rev. Lett. \textbf{104}, 106402 (2010).}

\bibitem{Cotlet16} O. Cotlet, S. Zeytinoglu, M. Sigrist, E. Demler and A. Imamoglu, \href{https://link.aps.org/doi/10.1103/PhysRevB.93.054510}{Phys. Rev. B \textbf{93}, 054510 (2016).}

\bibitem{Kavokin16} A. Kavokin and P. Lagoudakis, \href{http://dx.doi.org/10.1038/nmat4646}{Nature Mater. \textbf{15}, 599--600 (2016).}

\bibitem{Sentef18} M. A. Sentef, M. Ruggenthaler and A. Rubio, \href{https://arxiv.org/abs/1802.09437}{ArXiv: 1802.09437}

\bibitem{Fox01} M. Fox, \textit{Optical Properties of Solids}, (Oxford University Press, 2001).

\bibitem{Dressel02} M. Dressel and G. Gr\"{u}ner, \textit{Electrodynamics of Solids: Optical Properties of Electrons in Matter}, (Cambridge University Press, 2002).

\bibitem{Li18} X. Li et al., \href{https://doi.org/10.1038/s41566-018-0153-0}{Nature Photon. 1749-4893 (2018).}

\bibitem{Khveshchenko94} D. V. Khveshchenko and P. C. E. Stamp, \href{https://link.aps.org/doi/10.1103/PhysRevB.49.5227}{Phys. Rev. B \textbf{49}, 5227 (1994).}

\bibitem{Salam} A. Salam, \textit{Molecular Quantum Electrodynamics} (John Wiley \& Sons, 2010).

\bibitem{Scalapino86} D. J. Scalapino, E. Loh and J. E. Hirsch, \href{https://link.aps.org/doi/10.1103/PhysRevB.34.8190}{Phys. Rev. B \textbf{34}, 8190--8192 (1986).}

\bibitem{Romer15} A. T. R\o{}mer et al., \href{https://link.aps.org/doi/10.1103/PhysRevB.92.104505}{Phys. Rev. B \textbf{92}, 104505 (2015).}

\bibitem{Sigrist} V. P. Mineev and M. Sigrist, \textit{Basic Theory of Superconductivity in Metals Without Inversion Center}, \href{https://doi.org/10.1007/978-3-642-24624-1_4}{129--154 in: Non-Centrosymmetric Superconductors: Introduction and Overview (Springer, Berlin Germany 2012).}

\bibitem{Stemmer14} S. Stemmer and S. J. Allen, \href{https://doi.org/10.1146/annurev-matsci-070813-113552}{Ann. Rev. Mat. Res. \textbf{44}, 151 (2014).}

\bibitem{Manfra14} M. J. Manfra, \href{https://doi.org/10.1146/annurev-conmatphys-031113-133905}{Ann. Rev. Cond. Matt. Phys. \textbf{5}, 347 (2014).}

\bibitem{Damascelli03} A. Damascelli, Z. Hussain and Z.-X. Shen, \href{https://link.aps.org/doi/10.1103/RevModPhys.75.473}{Rev. Mod. Phys. \textbf{75}, 473--541 (2003).}

\bibitem{Hutchinson12} J. A. Hutchinson, T. Schwartz, C. Genet, E. Deveaux and T. W. Ebbesen, \href{http://dx.doi.org/10.1002/anie.201107033}{Angew. Chemie \textbf{51}, 1592--1596 (2012).}

\bibitem{Wang14} S. Wang et al., \href{http://dx.doi.org/10.1039/C4NR01971G}{Nanoscale \textbf{6}, 7243-7248 (2014).}

\bibitem{Orgiu15} E. Orgiu et al., \href{http://www.nature.com/nmat/journal/v14/n11/full/nmat4392.html}{Nature Mat. \textbf{14}, 1123 (2015).}

\bibitem{Ebbesen16} T. W. Ebbesen, \href{http://dx.doi.org/10.1021/acs.accounts.6b00295}{Acc. Chem. Res. \textbf{49}, 2403--2412 (2016).}

\bibitem{Feist15} J. Feist and F. J. Garcia-Vidal, \href{https://link.aps.org/doi/10.1103/PhysRevLett.114.196402}{Phys. Rev. Lett. \textbf{114}, 196402 (2015).}

\bibitem{Schachenmayer15} J. Schachenmayer, C. Genes, E. Tignone and G. Pupillo, \href{https://link.aps.org/doi/10.1103/PhysRevLett.114.196403}{Phys. Rev. Lett. \textbf{114}, 196403 (2015).}

\bibitem{Galego15} J. Galego, F. J. Garcia-Vidal and J. Feist, \href{https://link.aps.org/doi/10.1103/PhysRevX.5.041022}{Phys. Rev. X \textbf{5}, 041022 (2015).}

\bibitem{Kowalewski16} M. Kowalewski, K. Bennett and S. Mukamel, \href{http://dx.doi.org/10.1063/1.4941053}{J. Chem. Phys. \textbf{144}, 054309 (2016).}

\bibitem{Kowalewski16b} M. Kowalewski, K. Bennett and S. Mukamel, \href{http://dx.doi.org/10.1021/acs.jpclett.6b00864}{J. Phys. Chem. Lett. \textbf{7}, 2050-2054 (2016).}

\bibitem{Galego16} J. Galego, F. J. Garcia-Vidal and J. Feist, \href{http://dx.doi.org/10.1038/ncomms13841}{Nature Comm. \textbf{7}, 13841 (2016).}

\bibitem{Herrera16} F. Herrera and F. C. Spano, \href{https://link.aps.org/doi/10.1103/PhysRevLett.116.238301}{Phys. Rev. Lett. \textbf{116}, 238301 (2016).}

\bibitem{Galego17} J. Galego, F.J. Garcia-Vidal and J. Feist, \href{https://arxiv.org/abs/1704.07261}{arXiv: 1704.07261 (2017).}

\bibitem{Flick17} J. Flick, M. Ruggenthaler, H. Appel and A. Rubio, \href{http://www.pnas.org/content/114/12/3026.abstract}{Proc. Nat. Acad. Sci. \textbf{114}, 3026--3034 (2017).}

\bibitem{Flick18} J. Flick, C. Sch\"{a}fer, M. Ruggenthaler, H. Appel and A. Rubio, \href{https://doi.org/10.1021/acsphotonics.7b01279}{ACS Photonics \textbf{5}, 992 (2018).}

\bibitem{Woods16} L. M. Woods et al., \href{https://link.aps.org/doi/10.1103/RevModPhys.88.045003}{Rev. Mod. Phys. \textbf{88}, 045003 (2016).}


\end{thebibliography}
\end{document}